\magnification=\magstep 1
\hsize=134mm 
\vsize=200mm 
\hoffset=0mm 
\voffset=-2mm 
\parindent=0mm \parskip=7pt plus1pt minus.5pt 
\global\newcount\secno \global\secno=0
\def\newsec#1{\global\advance\secno by1 \bigbreak\mgap
{\bbf\the\secno. #1} $\hbox{\vrule width0pt height0pt depth13pt}$
\nobreak\newline \message{(\the\secno. #1)}} 
\global\newcount\subsecno \global\subsecno=0 
\def\newsubsec#1{\global\advance\subsecno by1
\bigbreak\sgap {\bf\the\secno.\the\subsecno. #1} $\hbox{\vrule
width0pt height0pt depth8pt}$ \nobreak\newline
\message{(\the\secno.\the\subsecno. #1)}} 
\global\newcount\equation \global\equation=0 
\def\nummer{\global\advance\equation by1 \eqno(\the\equation) }

\font\hbf=cmbx10 scaled\magstep2 
\font\bbf=cmbx10 scaled\magstep1 
\font\sbf=cmbx9
\font\srm=cmr9 
 
\font\caps=cmcsc10

\def\newline{\hfil\break\noindent}

\def\sgap{\par\vskip 3mm\noindent} 
\def\mgap{\par\vskip 5mm\noindent} 
\def\bgap{\par\vskip 8mm\noindent} 

\def\hh{{\cal H}}  \def\a{\alpha}  \def\b{\beta}  \def\udl{\underline}
\def\frac#1#2{{#1\over#2}}  \def\supp{{\rm supp}\,}  \def\id{{\rm id}}
\def\double#1{\,{}\colon #1 \colon\,}  
\def\trip{\vcenter{\vbox{\hbox{$.$}\hrule width.0ex height.45ex
                         \hbox{$.$}\hrule width.0ex height.45ex \hbox{$.$}}}}
\def\triple#1{\,\trip #1 \trip\,}  
\def\norm#1{\vert\!\vert #1 \vert\!\vert}
\def\be{$$} \def\ee{\nummer $$}
\def\RR{{\rm I\!R}} 
\def\cite#1{[#1]}
DESY 96-092 \hfill hep-th/9605156 \newline
May 1996
\mgap 
\centerline{\hbf Bounded Bose fields} 
\mgap
\centerline{\caps K.-H.\ Rehren} \sgap
\centerline{II.\ Institut f\"ur Theoretische Physik} 
\centerline{Universit\"at Hamburg ({\caps Germany})}
\centerline{{\srm email: rehren@x4u2.desy.de}}
\bgap 
\baselineskip12pt 
{\sbf Abstract.} 
{\srm Examples for bounded Bose fields in two dimensions are presented.}   
\baselineskip15pt 
\newsec{Introduction} 
Scalar free fields give rise to unbounded smeared field operators
$\varphi(f) = \int f(\udl x)\varphi(\udl x)\, d^dx$. In contrast \cite{1}, 
free Fermi 
fields give rise to bounded smeared field operators $\psi(f) = \int f(\udl x)
\psi(\udl x)\, d^dx$. These familiar facts are due to the unlimited occupation 
number in every excitation mode of the symmetric Fock space, and to the Pauli
exclusion principle in the anti-symmetric Fock space, respectively.

The boundedness or unboundedness being a probe for the structure of the state
space (in a sense suggested by the above examples), the question is of 
interest whether unboundedness of smeared field operators is a necessary 
feature of Bose fields. 
Indeed, Baumann \cite{2} has proven that the spectrum condition entails that 
{\it chiral}\/ Bose fields in two dimensions are necessarily unbounded. On the 
other hand, Buchholz \cite{3} has observed that a very simple Bose field 
in two dimensions which is the tensor product of two chiral free Fermi fields, 
that is
\be \varphi(f) = \int d^2x \, f(t,x) \, \psi(t+x) \otimes \psi(t-x) \; , 
\ee 
is bounded. Clearly, this field commutes with itself not only at space-like 
distance but also at time-like distance. In view of Huygens' principle, this 
feature is usually interpreted as the absence of interaction for massless 
fields. 

In the present letter we construct bounded Bose fields in two dimensions which 
commute with themselves at space-like distance but not at time-like distance.

All our examples are based on various factorizations of chiral Fermi 
fields into chiral vertex operators.

\newsec{The construction}
The starting point is the bosonization formula \cite{4}, which represents 
chiral free Fermi fields as vertex operators of unit charge. Using the
boundedness of smeared free Fermi fields, we shall infer the boundedness of 
smeared chiral vertex operators with charge below unity. In a second step, we 
pass to two-dimensional local fields.

We briefly recall the relevant definitions. Chiral vertex operators 
$E_\a(x)$ are defined by
\be 
   E_\a(x) = \mu^{\frac 12 \a^2} \triple{e^{i\a\phi(x)}} \equiv
   \mu^{\frac 12 \a^2} e^{i\a\phi_+(x)}e^{i\a\phi_-(x)} = E_{-\a}(x)^* 
\ee 
where $\phi = \phi_+ + \phi_-$ with $\phi_+ = (\phi_-)^*$, satisfying the
commutation relations
\be 
   [\phi_-(x),\phi_+(y)] = -\log i\mu(x-y-i\epsilon)  
\ee 
while $[\phi_-,\phi_-] = [\phi_+,\phi_+] = 0$. The variable $x \in \RR$ is a 
chiral light-cone coordinate. The regulating parameter $\mu > 0$ for the 
highly singular field $\phi$ will ultimately be sent to zero. Expressions 
involving $i\epsilon$ are always understood as boundary values as 
$\epsilon\searrow 0$.

The vertex operators (2) satisfy commutation relations
\be  
   E_\a(x) E_\b(y) = e^{\pm i\pi\a\b} E_\b(y)E_\a(x) 
\ee 
with the $+$ sign if $x>y$ and the $-$ sign if $x<y$.

By definition, the vacuum vector is annihilated by the negative frequency part,
\be  
   \phi_-\Omega = 0 \qquad \Leftrightarrow \qquad e^{i\a\phi_-}\Omega = 
   \Omega \; .
\ee 
This property determines the correlation functions among $E_\a(x)$:
\be
   (\Omega, E_{\a_1}(x_1) \cdots E_{\a_r}(x_r) \Omega) = 
   \mu^{\frac 12 (\sum \a_i)^2} \prod_{i<j} \Delta(x_i-x_j)^{-\a_i\a_j} 
\ee 
where
\be 
   \Delta(x) = \frac{-i}{x-i\epsilon} = \int_0^\infty dk e^{-ikx} \; . 
\ee 
In the limit $\mu\searrow 0$, the only correlations (6) which survive are 
those with total charge $\sum_i\a_i = 0$. By inspection of the correlation
functions, one verifies that in this limit
\be
   \psi(x) = E_{+1}(x) \qquad {\rm and} \qquad \psi^*(x) = E_{-1}(x) 
\ee 
are chiral free Fermi fields satisfying the canonical anti-commutation 
relations
\be
   \{\psi^*(x),\psi(y)\} = 2\pi\,\delta(x-y), 
   \quad\qquad \{\psi(x),\psi(y)\} = 0 \; .
\ee 
The derivative of the original field $\phi$ is related to the current 
$j = \double{\psi^*\psi}$ by
\be
   \partial \phi(x) = 2\pi j(x) \; . 
\ee 
The charge operator $Q = \int j(x)dx = \frac 1{2\pi} (\phi(\infty) - 
\phi(-\infty))$ measures the total charge $\sum_i \a_i$:
\be
   Q \; \prod E_{\a_i}(x_i)\Omega = 
   {\textstyle (\sum_i \a_i)} \; \prod E_{\a_i}(x_i)\Omega 
\ee 
provided $\mu\searrow 0$. Only in this limit, the vacuum is invariant under 
the $U(1)$ gauge symmetry $e^{itQ}$. 

Unlike the singular field $\phi$, the 
vertex operators $E_\a(x)$ are operator valued distributions on a Hilbert
space with positive metric. They are non-local conformally covariant fields 
with chiral scaling dimension $h = \frac 12 \a^2$.

All the previous holds true if one starts from a multi-component field
$\udl\phi$, that is
\be
   [\phi^a_-(x),\phi^b_+(y)] = -\delta^{ab} \log i\mu(x-y-i\epsilon) 
   \qquad (a,b = 1,\dots n) \; . 
\ee 
One only has to read quadratic expressions such as $\a^2$ or $\a\phi$ 
as scalar products $\udl\a^2 = \sum_a (\a^a)^2$ or $\udl\a\udl\phi = 
\sum_a \a^a \phi^a$, etc. $E_{\udl\a}$ is a chiral free Fermi field
whenever $\udl\a^2 = 1$, but different such fields do not anti-commute
with each other. They commute, actually, when $\udl\a'\udl\a = 0$.

\newsubsec{Factorization}
For our purpose, it is sufficient to consider vertex operators with 
two-component charges $\udl\a = (\a^1,\a^2) = \a^1\udl e^1 + \a^2\udl e^2$. 
Since $\phi^1$ and $\phi^2$ decouple, we may write
\be
   E_{\udl\a}(x) \equiv \mu^{\frac 12 \udl\a^2} \triple{e^{i\udl\a\udl\phi(x)}}
   = E_{\a^1\udl e^1}(x) \cdot E_{\a^2\udl e^2}(x) 
   \cong E_{\a^1}(x) \otimes E_{\a^2}(x) \; .
\ee 
Namely, the cyclic Hilbert space $\hh$ for all vertex operators with 
two-component charges decomposes as a tensor product of the cyclic Hilbert 
spaces $\hh^a$ for the vertex operators with charges $\a\udl e^a$ in the
basis directions ($a=1,2$), and the latter are isomorphic to vertex operators 
with scalar charges $\a$.
 
Note that the distributional meaning of the tensor product here is different 
from the one in the introduction: eq.\ (13) is the pointwise product of 
distributions in the same chiral variable which is well defined due to 
spectral (or analytic) properties of the correlation functions (6), while eq.\ 
(1) corresponds to the product of distributions in two independent chiral 
variables and yields a distribution over $\RR^2$.

Now we take $\udl\a^2 = 1$, $\udl\a = (\cos\tau,\sin\tau)$ such that $\psi 
\equiv E_{\udl\a}$ is a free Fermi field. The latter is not only defined on 
its own cyclic Hilbert space $\hh_\psi$ but on the much larger Hilbert space 
$\hh$ into which $\hh_\psi$ is embedded. 

We evaluate $\psi$ in a partial state $(\id \otimes \omega)$ where $\omega$ 
is some vector state $\omega = (\Phi, \cdot \, \Phi)$ of $\hh^2$. (For an 
operator $A$ on $\hh$, $(\id \otimes \omega)(A)$ is defined by its matrix 
elements $(\Psi_1 \otimes \Phi, A \Psi_2 \otimes \Phi)$.) We obtain an 
operator valued distribution on $\hh^1$:
\be
   (\id \otimes \omega)(\psi(x)) = E_{\cos\tau}(x) \cdot
   \omega(E_{\sin\tau}(x)) \; . 
\ee  
Solving for $\psi_\tau(x) \equiv E_{\cos\tau}(x)$ on $\hh^1$ and smearing with 
a test function $g$, we find that this field can be represented by the free
Fermi field evaluated in a partial state,
\be 
   \psi_\tau(g) = (\id \otimes \omega)(\psi(m_\omega g)) \; ,
\ee 
where $m_\omega(x) = \omega(E_{\sin\tau}(x))^{-1}$. We have to convince
ourselves that for an appropriate choice of the state $\omega$ this reciprocal
exists.

A convenient choice is given by the vector $\Phi = \Omega + \Omega_{\sin\tau}$ 
where $\Omega_{\sin\tau}$ is the ground state for the conformal Hamiltonian 
$L_0$ (with `energy' $h = \frac 12 \sin^2\tau$) within the sector of total 
charge $Q = \sin\tau$. With this choice, $\omega(E_{\sin\tau}(x)) = 
(\Omega_{\sin\tau},E_{\sin\tau}(x) \Omega) \sim (1-ix)^{-\sin^2\tau}$, so the 
reciprocal function 
\be 
   m_\omega(x) \equiv \omega(E_{\sin\tau}(x))^{-1} 
   = const. \cdot (1-ix)^{\sin^2\tau} 
\ee 
is well defined.

\newsubsec{Boundedness}
Chiral free Fermi fields, and hence the vertex operators with charge 
$\udl\a^2 = 1$ satisfy the $L^2$-bound 
\be 
   \norm{\psi(f)} = \norm{E_{\udl\a}(f)} \leq const. \cdot \norm{f}_2 \; . 
\ee 
This bound is algebraically determined \cite{1} by the canonical 
anti-commutation relations (9), and therefore holds not only on the cyclic 
Hilbert space $\hh_\psi$ but on the entire Hilbert space $\hh$.

It is elementary to show that a partial state $(\id \otimes \omega)$
takes bounded operators on $\hh^1 \otimes \hh^2$ into bounded operators
on $\hh^1$,
\be 
   \norm{(\id \otimes \omega)(A)} \leq \norm{\omega} \; \norm{A} \; .
\ee 
Hence by eq.\ (15), $\psi_\tau(g)$ is bounded by
\be 
   \norm{\psi_\tau(g)} \leq \norm{\omega} \; \norm{\psi(m_\omega g)} \leq 
   const. \cdot \norm{m_\omega g}_2 
\ee 
whenever the test function $g$ is in the domain of the multiplication operator
$(m_\omega g)(x) = m_\omega(x) g(x)$.

For the choice of $\omega$ as above, $m_\omega$ has dense domain in 
$L^2(\RR)$. Clearly, all Schwartz functions are in the domain of $m_\omega$. 
Hence we have established the $L^2$-bound (19) for vertex operators with 
charge below unity smeared with any test function of sufficiently rapid decay.

We note that the multiplication operator $m_\omega$ is absent in the 
`compact' description of chiral fields. Namely, the conformal vertex 
operators $E^c_\a(z)$ on the compactified light-cone $S^1$ (where 
$z = \frac{1+ix}{1-ix} \in S^1$) given by
\be  
   E^c_\a(z) = (dz/dx)^{-\frac 12 \a^2} E_\a(x) 
   = const. \cdot (1-ix)^{\a^2} E_\a(x)
\ee 
acquire a weight factor $\sim (1-ix)^{\a^2}$. The different weight factors 
pertaining to $\psi^c_\tau$ (with $\a^2 = \cos^2\tau)$) and $\psi^c$ (with 
$\udl\a^2 = 1$), respectively, precisely cancel the multiplication operator 
$m_\omega$ given by (16), and one has $\psi^c_\tau(z) = const. \cdot 
(\id \otimes \omega)(\psi^c(z))$. Thus, $\psi^c_\tau$ satisfies the same 
$L^2(S^1)$-bound as the chiral free Fermi field on the circle.

\newsubsec{Locality of tensor products}
Due to eq.\ (4), the bounded operators $\psi_\tau(g)$ satisfy the non-local 
commutation relations
\be 
   \psi_\tau(g_1)\psi_\tau(g_2) = e^{\pm i\pi\cos^2\tau} \psi_\tau(g_2)
   \psi_\tau(g_1) 
\ee 
whenever $\supp g_1 > \supp g_2$ resp.\ $\supp g_2 > \supp g_1$. 

In order to obtain Bose fields, we proceed to the tensor product of two 
non-local chiral fields $\psi_\tau$
\be 
   \varphi_\tau(g \otimes h) := \psi_\tau(g) \otimes \psi_\tau(h) 
\ee 
for test functions $(g \otimes h)(t,x) = g(t+x)h(t-x)$ on $\RR^2$. This 
tensor product is understood to extend by linearity, as in eq.\ (1), to the 
two-dimensional field
\be \varphi_\tau(f) = 
\int d^2x \, f(t,x) \, \psi_\tau(t+x) \otimes \psi_\tau(t-x) \; . 
\ee 

If two test functions $f_i = g_i \otimes h_i$ on $\RR^2$ ($i = 1,2$) have 
supports at space-like distance, then either $\supp g_1 < \supp g_2$ and 
$\supp h_1 > \supp h_2$, or $\supp g_1 > \supp g_2$ and $\supp h_1 < 
\supp h_2$. In either case, the commutation relations (21) produce two
opposite phase factors, and consequently $\varphi_\tau(f_1)$ and 
$\varphi_\tau(f_2)$ commute, whenever $f_i$ have supports at space-like 
distance. On the other hand, if $f_i$ have supports at time-like distance, 
then the phase factors add rather than cancel, and consequently 
$\varphi_\tau$ does not satisfy Huygens' principle. Choosing $\cos^2\tau = 
\frac 12$ one may even have time-like anti-commutation. (This possibility
was also previously discovered by Buchholz \cite{3}.)

By linearity, the local commutativity extends to tensor product fields 
smeared with arbitrary test functions on $\RR^2$. 

\newsubsec{Boundedness of tensor products}
By Schwartz' nuclear theorem, the extension of the tensor product of 
two (operator-valued) distributions in one variable to a functional in two 
variables is again a distribution, and the chiral bound (19) implies bounds 
for the two-dimensional fields (23) in terms of some Schwartz norms of the
test function $f$. The following estimates due to Buchholz \cite{3} improve 
the bounds due to the theorem.

We rewrite eq.\ (23) in the form
\be \varphi_\tau(f) = \int dp\,dq\;P(i\partial_p)P(i\partial_q)\tilde f(p,q)
\; \int d^2x \frac{e^{ipx_+}}{P(x_+)} \psi_\tau(x_+) \otimes 
\frac{e^{iqx_-}}{P(x_-)} \psi_\tau(x_-) \; ,
\ee 
where $\tilde f(p,q)$ is the Fourier transform of $f$ with respect to the 
chiral coordinates $x_\pm = t \pm x$, and $P$ is a polynomial in one chiral 
coordinate such that $1/P$ is in the domain of $m_\omega$. The 
operator-valued integral $\int d^2x \ldots$ in eq.\ (24) is of the form (22), 
and is hence of norm less than $const. \cdot \norm{m_\omega/P}_2^2$. Thus
\be \norm{\varphi_\tau(f)} \leq const. \cdot
\norm{P(i\partial_p)P(i\partial_q)\tilde f}_1 \; .
\ee 
Estimating the $L^1$-norm by the Cauchy-Schwarz inequality
$\norm f_1 \leq \norm{1/Q}_2\norm{Qf}_2$ provided $1/Q$ is square integrable, 
and using the Plancherel equality of the $L^2$-norms of a function and its
Fourier transform, we also have
\be \norm{\varphi_\tau(f)} \leq const. \cdot
\norm{Q \cdot P(i\partial_p)P(i\partial_q)\tilde f}_2 = const. \cdot
\norm{Q(-i\partial_+,-i\partial_-) (P \otimes P) f}_2
\ee 
for appropriate polynomials $P(x_\pm)$ and $Q(p,q)$ as 
qualified before, e.g., $P(x) = 1+x^2$ and $Q(p,q) = (1-ip)(1-iq)$. The
bound (26) holds for all test functions $f$ which are sufficiently smooth
and of sufficiently rapid decay, and in particular for all Schwartz functions. 
The constant depends on $\tau$, on the state $\omega$, and on $P$ and $Q$.

We conclude that $\varphi_\tau$ are bounded Bose fields with non-trivial 
time-like commutators. As the parameter $\tau$ varies between $0$ and 
$\frac \pi 2$, the fields $\varphi_\tau$ interpolate between Buchholz' 
example (1) and the identity operator.

\newsubsec{Further examples}
The above construction has a non-abelian generalization. It was shown by
Wassermann and by Loke, with essentially the same method as ours, that -- 
when passing from level $k$ to level $k+1$ -- the coset model factorization 
of chiral free $SU(N)$ Fermi fields produces $L^2$-{\it bounded}\/ chiral 
exchange fields (non-abelian vertex operators) 
which are primary for the level $k+1$ current algebra \cite{5}, and for the 
coset Virasoro algebra (for $N=2$) \cite{6}, respectively. The $SU(2)$ coset 
Virasoro algebra has central charge in the discrete series $c<1$, and the 
primary chiral fields thus obtained are, in the standard nomenclature,
the fields $\phi_{12}$ and $\phi_{22}$ with chiral scaling dimension 
$h_{12} = \frac{k-1}{4(k+2)}$ and $h_{22} = \frac 3{4(k+1)(k+2)}$. 
The primary fields for the current algebra are those corresponding to the
vector and conjugate vector representations of $SU(N)$.

On the other hand, it is well known (see, e.g., \cite{7}) that tensor 
products of chiral exchange fields in rational theories yield {\it local}\/ 
two-dimensional fields of the form
\be 
   \varphi(t,x) = \sum_{e\bar e} \zeta_{e\bar e} \; \phi_e(t+x) \otimes 
   \phi_{\bar e}(t-x) 
\ee 
with appropriate numerical coefficients $\zeta_{e\bar e}$; the labels $e$
and $\bar e$ run over the pairs of initial and final sectors connected by
the primary field $\phi$ and its conjugate, respectively. This result relies 
only on a `CPT' symmetry due to pentagon and hexagon identities of fusion and 
braiding matrices.

Taken together, these two results show that $\varphi$ in eq.\ (27) are bounded 
two-dimensional Bose fields if the primary field $\phi$ is one of the fields
produced by the coset factorization of chiral free Fermi fields. 
These fields are the fields $\varphi_{12}$ and $\varphi_{22}$ in the minimal 
models \cite{8}, and the basic matrix fields $g_{ab}$ of the two-dimensional 
$SU(N)$ Wess-Zumino-Witten models \cite{9}, respectively.

\newsec{Conclusion}
We have presented a variety of bounded Bose fields in two dimensions. The
family of fields (23) for various parameters $\tau$ are actually defined 
on a common Hilbert space and are relatively local among each other. 

All the bounded Bose fields we have constructed here are conformally covariant
fields with scaling dimension $d \equiv h_+ + h_- \leq 1$. To remove this 
upper bound for the scaling dimension, one may consider (non-primary) 
derivative fields
\be 
   \partial_{\mu_r} \ldots \partial_{\mu_1}\varphi(f) = (-1)^r
   \varphi(\partial_{\mu_r} \ldots \partial_{\mu_1}f) 
\ee 
which are again bounded Bose fields but with scaling dimensions
$r \leq d \leq r+1$.

\bigbreak

We conclude that in two dimensions, bounded Bose fields are quite abundant. 
We still do not know whether bounded Bose fields exist in more than
two dimensions.
\mgap
 
{\bf Acknowledgments.} I thank J. Yngvason for encouraging me to write 
down these notes, and D. Buchholz and K. Fredenhagen for helpful comments.

\mgap

{\bbf References} \vskip 3mm
\def\ref#1{\par \noindent \hangafter=1 \hangindent 14pt \cite{#1}}
\parskip 4pt
\baselineskip=2.6ex\smallskip
\def\CMP#1{Com\-mun.\ Math.\ Phys.\ {\bf #1}}
\def\HPA#1{Helv.\ Phys.\ Acta {\bf #1}}
\def\NP#1{Nucl.\ Phys.\ {\bf #1}}
\def\PR#1{Phys.\ Rev.\ {\bf #1}}
\ref{1} H. Araki, W. Wyss: {\it Representations of canonical 
anticommutation relations}, \HPA{37}, 136--159 (1964). 
\ref{2} K. Baumann: Talk at the Symposium on Algebraic Quantum Field Theory
and Constructive Field Theory, G\"ottingen, 1995, and preprint in 
preparation (1996).
\ref{3} D. Buchholz: unpublished.
\ref{4} S. Mandelstam: {\it Soliton operators for the quantized sine-Gordon 
equation}, \PR{D 11}, 3026--3030 (1975).
\ref{5} A. Wassermann: {\it Operator algebras and conformal field theory,
III}, preprint Cambridge (UK), 1995.
\ref{6} T. Loke: {\it Operator algebras and conformal field theory of the 
discrete series representations of {\rm Diff}$(S^1)$}, PhD thesis Cambridge 
(UK), 1994.
\ref{7} K.-H. Rehren: {\it Space-time fields and exchange fields}, \CMP{132},
461--483 (1990).
\ref{8} A.A. Belavin, A.M. Polyakov, A.B. Zamolodchikov: {\it Infinite
dimensional symmetries in two-dimensional quantum field theory}, 
\NP{B 241}, 333--380 (1984).
\ref{9} V.G. Knizhnik, A.B. Zamolodchikov: {\it Current algebra and 
Wess-Zumino model in two dimensions}, Nucl.\ Phys.\ {\bf B 247}, 83--103
(1984).
\bye